\documentclass[a4paper,12pt]{article}
\usepackage{natbib}
\usepackage{graphicx,subfig}
\usepackage{amsmath}
\usepackage{amssymb}
\usepackage{hyperref}
\usepackage{xcolor}
\begin{document}
\title{Solar supergranular fractal dimension dependence on the Solar cycle phase}
\author{Rajani G$^1$, Sowmya G M$^2$, U Paniveni$^{3,4}$, R. Srikanth$^4$ \\
$^1$ PES College of Engineering, Mandya-571401,Karnataka,India. \\
$^2$ GSSS Institute of Engineering and Technology for Women, KRS Road,\\Metagalli Mysuru-570016, Karnataka, India.\\
$^3$ Bangalore University, Jnanabharathi, Bengaluru – 560056.\\
$^4$ Poornaprajna Institute of Scientific Research, Devanahalli, \\
Bangalore-562110, Karnataka, India.
}
\date{}
\maketitle

\begin{abstract}
We study the complexity of the supergranular network through fractal dimension by using Ca II K   digitized data archive obtained from Kodaikanal solar observatory. The data consists of 326 visually selected supergranular cells spread across the 23rd solar cycle. Only cells that were well-defined were chosen for the analysis and we discuss the potential selection effect thereof, mainly that it favors cells of a smaller size ($< 20$ Mm). Within this sample, we analyzed the fractal dimension of supergranules across the Solar cycle and find that it is anticorrelated with the activity level. 
%Our study supports the view that supergranules correspond to large-scale outflows of the solar convection and that magnetic confinement may play a role in regularizing cell boundaries.
\end{abstract}

Keywords : Sun: photosphere, Sun : granulation, stars: activity, Sun : magnetic  fields

\section{INTRODUCTION}\label{introduction}
Solar convection occurs over a continuum of scales, mainly around 1 Mm, corresponding to granulation, and around 30 Mm, corresponding to supergranulation \citep{hathaway2000photospheric, rieutord2010sun}. \citet{meunier2007intensity} show a low-level intensity contrast and infer a corresponding temperature difference of 0.8K-2.8 K between the center of the cells and boundaries, indicative of a convective origin of supergranulation. As regards the spread in scale, the kinetic energy spectrum having a distinct peak at wavelengths of $\sim$ 35 Mm consists of  cells at least three times larger and extends to much smaller cells of scales traditionally associated with granulation\citep{hathaway2000photospheric}. 

However, depending on the method chosen, other comparable scales for the latter have been reported. For example, \citet{rieutord2008solar} find supergranulation spectrally extending between scales 20 Mm and 75 Mm, while peaking at 36 Mm. \citet{chatterjee2017variation}, using data from the Kodaikanal Solar Observatory\footnote{https://kso.iiap.res.in} (KSO) similar to the one used in this study, report an average scale in the range of 22--28 Mm. Using Dopplergrams obtained from SOHO mission and Ca-K filter-grams,\citep{srikanth1999chromospheric} using a tessellation technique report a  supergranular mean scale of 25 Mm, with the kurtosis and skewness of the supergranular scale distribution being 4.6 and 1.1 respectively, in agreement with values derived by other methods (reported in the above paper). 

%Supergranulation, first reported by \citet{leighton1962velocity}, found the existence of large cellular patterns with a spacing of about 30 Mm between the centres of the adjacent cells in Dopplergrams and Ca-K line images of the sun.\citep{simon1964velocity}, using the auto-correlation function and cross-correlation function, showed that the mean cell size of the supergranule is 32 Mm. 

Interestingly, in contrast to the above estimates for cell size, \citep{parnell2009power} have obtained a characteristic cell diameter in the range 13-18 Mm, almost half the traditional cell sizes quoted above, by following the boundaries of individual cells using a method of tesselating the Solar surface. Here, we may also note that \citep{de2000near} and \citep{derosa2004evolution}, using local correlation tracking, derived a rather small diameter in the range 12-20 Mm. Also, the average cell size is 17.06 Mm at cycle minimum and 16.11 at cycle maximum as observed by \citep{meunier2008supergranules} using the granule tracking method. 

The cycle dependence remarked above is in agreement with the result of \citep{singh1981dependence}, who showed that size of the cells are smaller by about 5\% during the active phase of the solar cycle in comparison to that at the quiet phase. This is consistent with the idea that network magnetic elements having a shrinking effect on supergranules, as hinted by \citep{meunier2008supergranules}. However, a difference in supergranular scales for active and quiet regions of the Sun, the former being about 1.5 Mm larger, has been noted \citep{mandal2017association}. It is possible that some of the discrepancy in the reported data about supergranular scales and their properties is due to a potential bias in studies based on tracers such as Ca II K for indicating supergranulation \citep{rincon2018sun}. A latitude-dependence of supergranular scales, with a variation as large as 7\%, has been noted\citep{raju1998dependence}. Supergranules are generally known to live for about 30 hours or more., with larger cells living longer  \citep{hirzberger2008structure}. 

Supergranulation has dynamical interactions with the magnetic fields of the quiet sun and  most notably, supergranules are strongly correlated with the magnetic network \citep{rieutord2010sun}.  Solar supergranular convection is understood to play a major role in the global and local structuration and dynamics of solar magnetic ﬁelds at the interface between the solar interior and corona \citep{rincon2018sun}.

Fractal dimension analysis is a powerful mathematical tool to analyze the shape complexity of geometrical structures, by quantifying the degree of self-similarity of a set \citet{cannon1984fractal}. The fractal dimension of the supergranulation can give an indication of the turbulence in the Solar magneto-convection. A fractal dimension $D=1$ (resp., $D>1$) indicates cells that are regular (resp., more frazzled and hence space-filling). 

Fractal analysis in the context of a solar surface phenomenon was by \citet{muller1987dynamics} who reported a fractal dimension $D=1.25$ for smaller granules and $D \approx 2$ for larger ones. \citet{paniveni2005fractal} obtained a fractal dimension of about $D=1.24$ for supergranules found in the SOHO dopplergrams, which is closer to the dimension for smaller granules. Building on the latter result, \citep{paniveni2011solar,paniveni2015supergranulation} studied the turbulence in Solar magneto-convection. 

By Kolmogorov hypothesis in the context of a turbulent convection, the horizontal velocity $V_h$ of a convective cell varies with scale $L$ as $V_h \propto L^{1/3}$ \citep{krishan2002relationship}. A direct relation between cell fractal dimension and magnetic activity was reported by \citet{nesme1996fractal}, and similarly by \citep{meunier1999fractal} using both full disk and high-resolution MDI magnetograms. 

 While fractal dimension of active region supergranulation is consistently lower than that of quiet region supergranulation as reported by \citet{chatterjee2017variation}, the latter cells can have higher fractal dimension during the active phase of the Solar cycle, and the former cells contrariwise. It would appear that the full interplay of global features such as magnetic activity and Solar cycle phase with the supergranular outflow must be borne in mind to fully unravel the behavior of fractal dimension, scale or any other cell parameter.  
 
 Here, a pertinent question would be whether cell scale has a role in affecting its fractal dimension across the Solar cycle, which is addressed in part in this work. To this end, we have adopted the method of cell study used by \citep{paniveni2005fractal}, which is preferential towards smaller cells. Our analysis, restricted to the quiet region, suggests that while the cross-cycle behaviour remains qualitatively the same, smaller cells have larger than average fractal dimension.

\section{DATA and ANALYSIS}
The 23rd solar cycle (1996-2008) at  the Kodaikanal Solar Observatory has been used for this analysis. The Kodaikanal solar tower dual telescope houses a K-line spectroheliograph which is a 2-prism instrument along with 7{\AA} /mm spectral dispersion near 3930{\AA}. It works with a 60 mm image formed from a 30 cm Cooke photovisual triplet. 
A Foucault siderostat with 46 cm diameter reflects sunlight onto the 30 cm lens. Exit slits are centered at K 232 admits 0.5 A. By using the Photo digitizing system, the images are digitized in terms of strips which are running parallel to the equator. The resolution of the patterns obtained are of 2 arcsec which is two times the granular scale. Further, the data is time averaged over an interval of 10 min which is double the 5 min period of oscillation. The signal due to granular velocity is largely averaged out by time averaging and spatial resolution. Similarly, the contributions to p-mode vibrations are minimized after time averaging.

Cycle 23 lasted during the period  August 1996 -- December 2008.  The degree of activity determines the level of dispersal of magnetic fields, which is expected to influence properties of the cell network, in particular to lower the fractal dimension in quiet regions. Ideally, we expect a continuous transition of this property across the cycle. For our data, this behavior can be broadly captured by dividing the cycle into three phases-- peak, minimum and intermediate. The maximum or peak phase is identified with the period 2000--2002, during which large sunspots as well as other manifestations of activity, such as coronal mass ejections and Solar flares, are found to occur. The minimum phase is identified with the periods 1996--1997 and 2005--2007 during which such magnetic activity is minimal. The intermediate phase is identified with the intervening ascending and descending phases of the solar cycle. The threefold division here is a reasonably robust tradeoff between statistical significance of a phase period and the number of phase periods across the cycle.

	An initial selection of frames suitable to estimate the Area and Perimeter of supergranular cells are made. Regions in the quiet Sun, where the supergranules appear to be well pronounced, are delineated by visual inspection. Out of these regions, we selected well defined supergranular cells between $15^{\circ} < \theta < 30^{\circ}$, $\theta$ being the angular distance of the cell from the disc centre. Cells below the lower limit show diminished supergranular flow signature. The upper limit is chosen to keep the foreshortening effect low. The quiet region cells were individually selected, extracted and studied  as shown in Fig 1. in a manner similar to that applied to a study of dopplergrams \citep{paniveni2004relationship,paniveni2010activity}.\\
	\begin{figure}
  \subfloat[Full disk Ca II K filtergram]
		  {
			\begin{minipage}[c][1\width]{
					0.5\textwidth}
				\centering
				\includegraphics[width=1\textwidth]{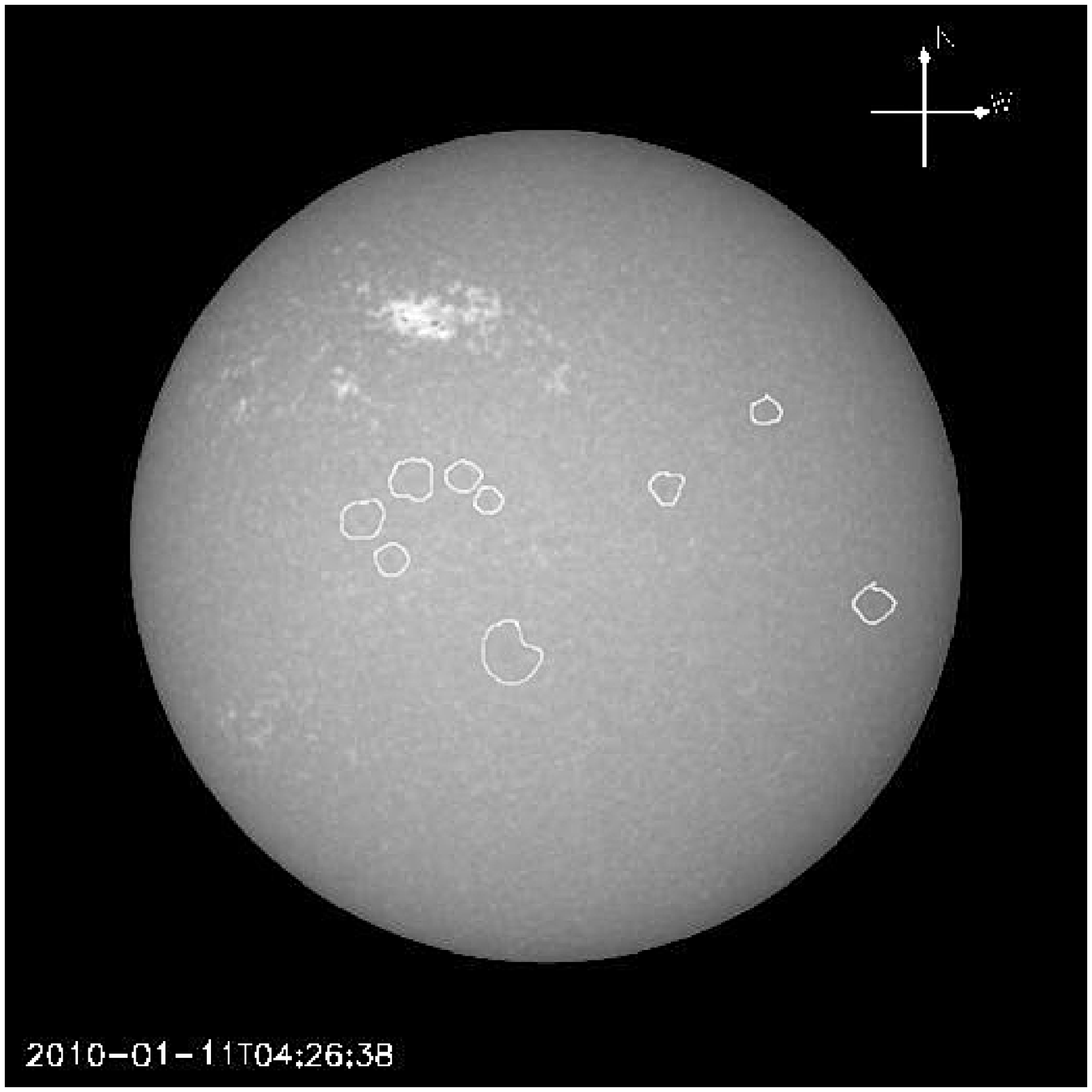}
		\end{minipage}}
\hfill
	\subfloat[Close-up of selected region]
		{
			\begin{minipage}[c][1\width]{
					0.5\textwidth}
				\centering
				\includegraphics[width=1\textwidth]{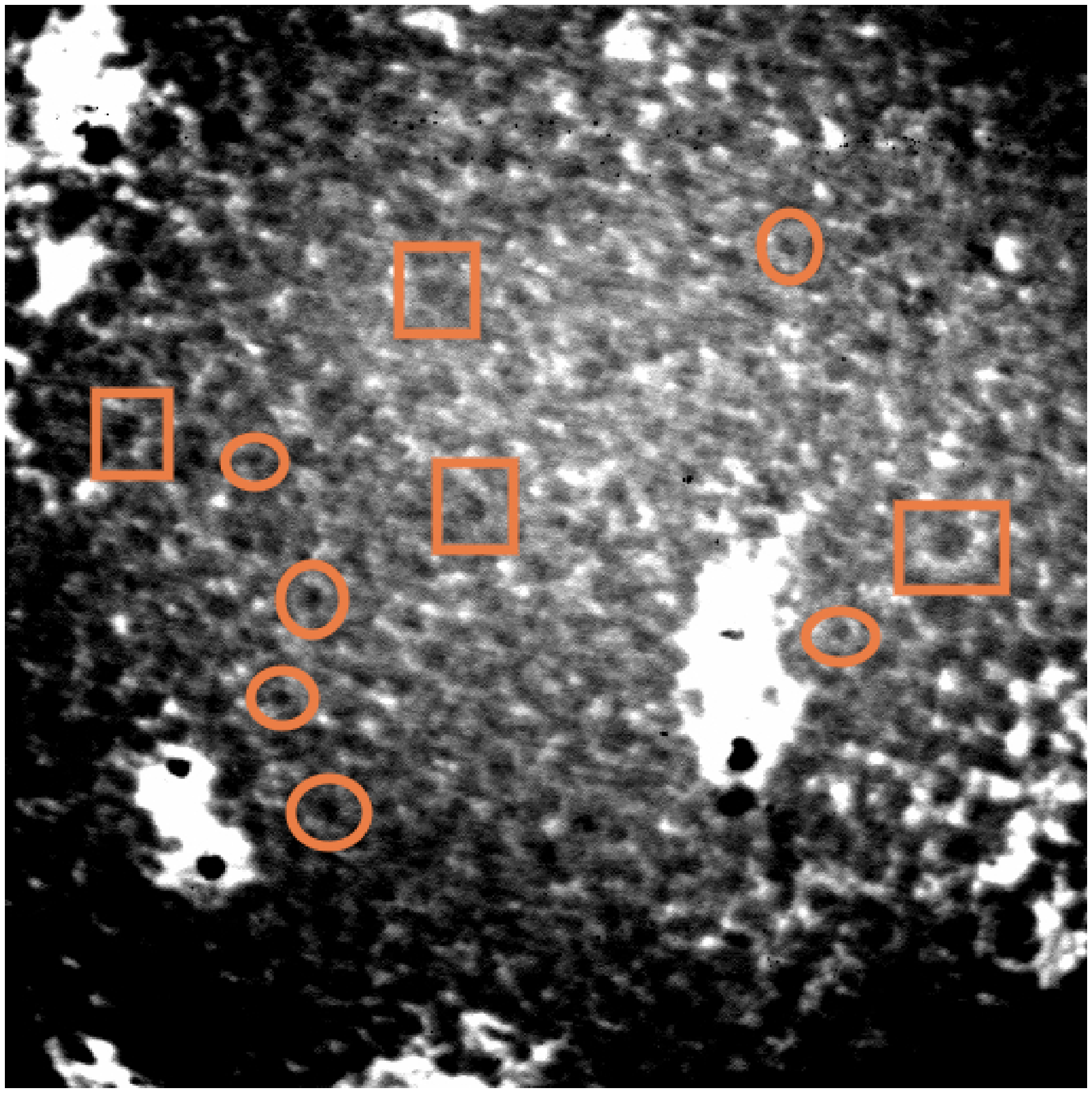}
		\end{minipage}}
		\caption{Data and cell selection: (a) A full disk Ca II K filtergram data of Kodaikanal Solar observatory where supergranules with well-defined boundaries are encircled. (b) Closeup of a selected region, where the circles enclose cells that are well-defined, and hence suitable for inclusion in this study, whilst the rectangles enclose cells that aren't so, and hence rejected. }
		\label{fig:KSO_fulldisc}
	\end{figure}  

The steps involved in detecting and extracting the supergranules from KSO intensity images are as follows. A typical full disk image that is the basis of our data is shown in Fig.~\ref{fig:KSO_fulldisc}(a), where a few potential cells for our analysis are encircled. In  Fig. \ref{fig:KSO_fulldisc}(b), we indicate supergranular cell that are eventually selected by a circle, and those that are eventually rejected by square enclosures.  

Cell selection is based on the requirement that the cells are well accentuated. Those so obtained are used to determine the area and perimeter for a given cell, and hence the spectrum for all selected supergranules. The area-perimeter relation is used to derive the fractal dimension \citep{paniveni2005fractal}. We analysed 326 reasonably well-defined cells identified in the quiet region, which is characterized by a low sunspot number, less solar irradiance and flare emissions hence by a weak magnetic field.

\begin{figure}
	\centering
	%	\begin{subfigure}{0.7\textwidth}
	\includegraphics[width=0.5\textwidth,angle=90]{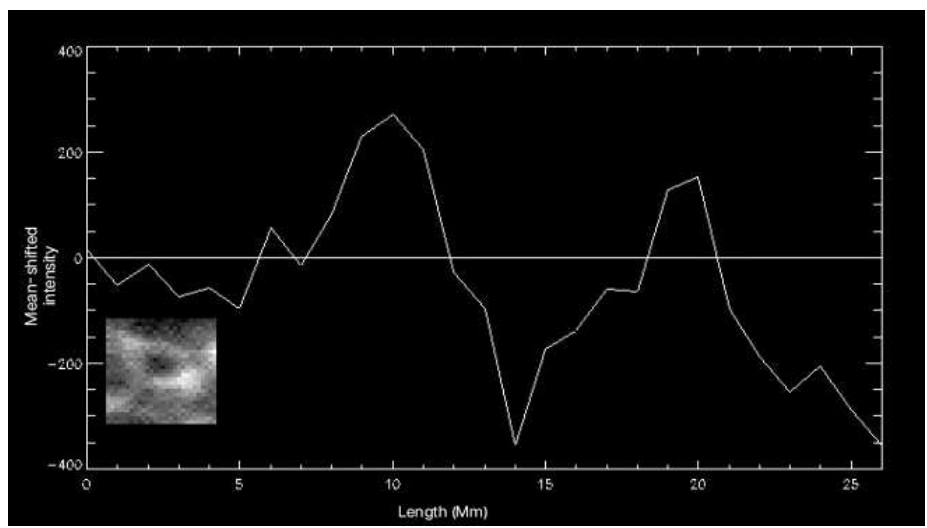} % png
	%	\end{subfigure}
	%	\begin{subfigure}{0.2\textwidth}
	%		\includegraphics[width=\textwidth]{a.eps} % png
	%	\end{subfigure}
	\caption{Intensity profile of a selected cell in Ca II K filtergram (inset): the width of the cell at this longitudinal location is the distance between the two peaks, representing the cell walls. The cell size is derived using the full set of such transverse scans across the cell. (Note: For convenience of display in terms of contrasting peaks and valleys in the intensity topography, the intensity values are mean-shifted, so that those smaller than the mean intensity are negative-valued.)}
	\label{fig:intensity_scan}	
\end{figure}

The profile of the visually identified cell was scanned transversely over the whole length: typically, we choose a fiducial y-direction on the cell and perform intensity profile scans along the x-direction for all the pixel positions on the y-axis (Fig. \ref{fig:intensity_scan}). In each scan, the cell extent is taken as the sum of the distance between the two consecutive, expected peaks in the intensitygram. The sum of these values at all y-levels multiplied by the pixel dimension gives the area of the cell.  Interactive data Language (IDL) is used for data visualization and analysis to derive area, perimeter and other parameters of the selected supergranular cell using this data.

This may be contrasted by other methods employed to study supergranulation, some even of the present authors in their previous works, such as cross-correlation, tessellation, etc, which are typically automated and allow handling large regions of supergranulation at once. The manual method we opt for here is thus biased towards cells are well defined in a qualitative sense, implying a rejection rate by area of about $\frac{4}{5}$. Later we discuss potential effects of this selective analysis, and argue that it is consistent with our fractal analysis.

Since the cell wall is formed by a heating of the overlying plasma by the magnetic flux swept by the supergranular convective flow, larger cells typically show more fluctuations and discontinuities in the cell wall. Thus, our requirement for well defined cells creates a selection effect towards cells that are smaller than the size scales obtained by other methods, about which we discuss later.

\section{RESULTS}
\begin{figure}
	\centering
	\includegraphics[width=0.85\textwidth]{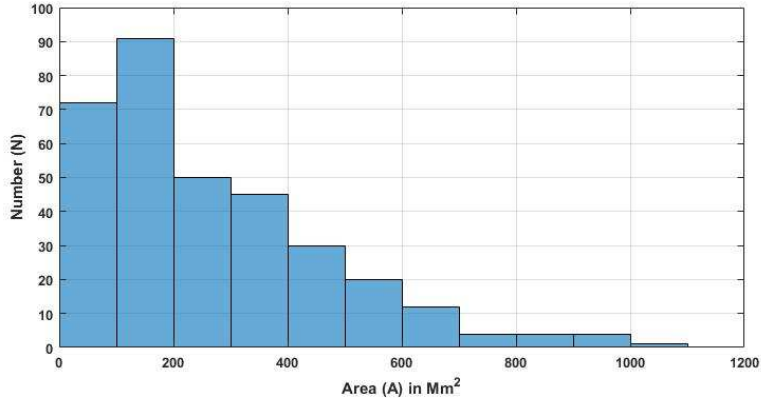}
	\caption{Histogram for the area of supergranular cells of the present data set. All cells belong to the 23rd Solar cycle based on the KSO data.}
	\label{fig:histo}
\end{figure}

\begin{figure}
	\centering
	\includegraphics[width=0.85\textwidth]{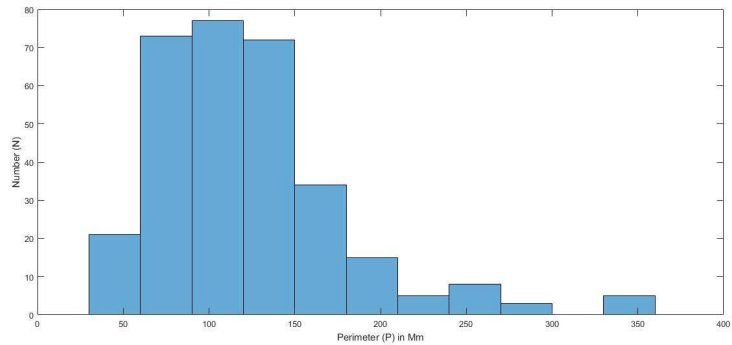}
	\caption{Histogram for the perimeter of supergranular cells of the present data set. All cells belong to the 23rd Solar cycle based on the KSO data.}
	\label{fig:histo_P}
\end{figure}

The basic statistical parameters concerning the maximum, mean, standard deviation for supergranular cell of area A and perimeter P for the 23rd solar cycle (1996--2007) are 1082, 269 and 199.7$\pm$ 11  for Area in Mm$^2$, and 357, 98.9 and 46.2$\pm$ 2.5  for perimeter in Mm respectively. A histogram of the area and perimeter data is given in Figures \ref{fig:histo} and \ref{fig:histo_P}, respectively.  

Skewness of a distribution is a measure of its asymmetry, and may  be either a positive value, negative value or zero corresponding to a right-skewed, left-skewed or symmetric distribution \citep{paniveni2005fractal}). For our present data, the skewness obtained is 1.2 for the area distribution, and 0.8 for Perimeter, indicating asymmetry with a bias towards larger values.
Kurtosis quantifies the clustering of distribution towards the center relative to the tails, and thus gives a measure of how peaked a distribution is. For our data, the values of kurtosis obtained are 1.4 and 0.6 for the area and perimeter distributions.  

%The mean cell area, when packed into a circle, corresponds to a length scale (diameter) of about 18.5 Mm, which is about 60\% to 75\% of the typical length scales reported (as cited in the Introduction). On the other hand, the mean perimeter we obtain, if assumed to border a circular cell, would correspond to a length scale of about 32 Mm, closer to usually reported values.

\begin{figure}[htp]
	\subfloat[Minimum phase]{
		\includegraphics[width=5in]{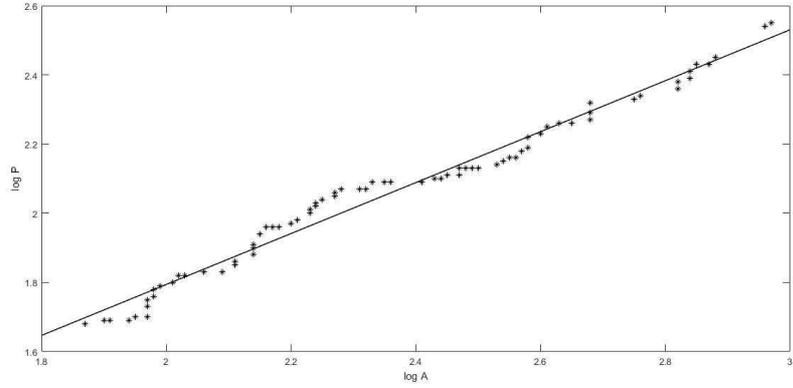}%
	}

	\subfloat[Intermediate phase]{%
		\includegraphics[width=5in]{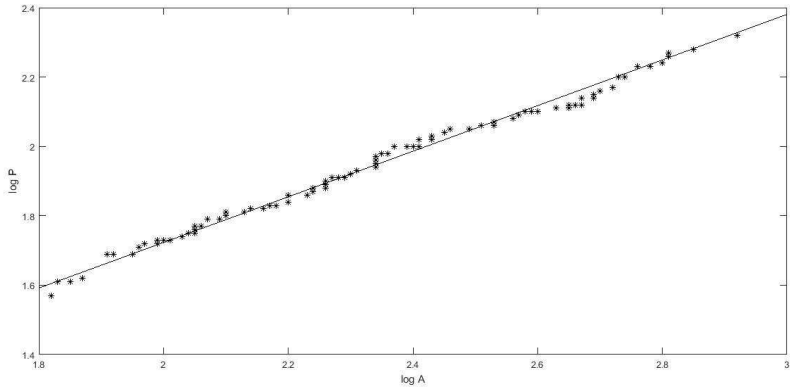}%
	}

	\subfloat[Peak phase]{%
	\includegraphics[width=5in]{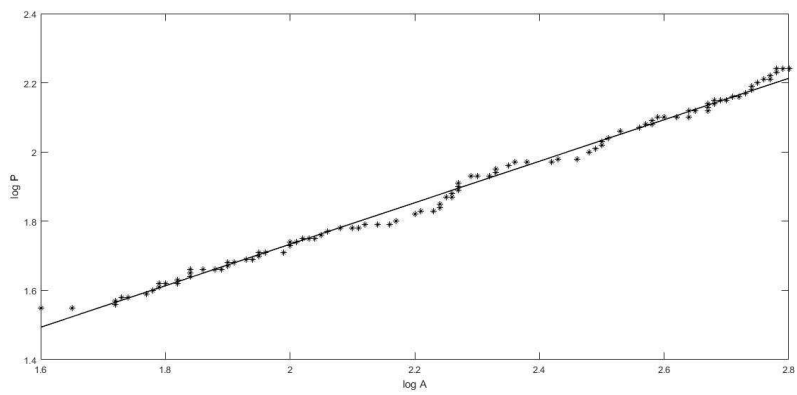}%
}
	\caption{Log-log plot of supergranular perimeter (in Mm) against supergranular area (in Mm$^2$) obtained for the (a) cycle minimum, (b) intermediate (middle) and (c) peak phases (base). The fractal dimensions read off the slope of these plots are summarized in Table \ref{tab:phases}.}
	\label{fig:A-P}
\end{figure}

The fractal dimension is generally found to be greater than 1, indicating that the boundary of the cell is not smooth but craggy, causing greater boundary length to enclose the same area. The degree to which the fractal dimension exceeds 1 is a measure of how rugged the cell boundary is.
The data on area and perimeter of the cells are plotted in Figures . The linearity in the log-log relation is quite apparent, indicating a power-law relation. Fig.~\ref{fig:A-P} is then used to read-off the fractal dimension $D$ (Hausdorff dimension) of the solar convection cells according to the relationship:
\begin{equation}
P=cA^{D/2}, 
\label{eq:A-P}
\end{equation} 
where P is perimeter,  C is constant, A is Area of the cell  and D is fractal dimension.

In our data, the area and perimeter are quite well correlated, with an average correlation coefficient of  0.99. Fractal dimension $D$  is calculated as 2 $\times$ slope and is found to be 1.2$\pm$ 0.06 using Eq. (\ref{eq:A-P}). 

Our results are comparable to that reported by \citet{paniveni2005fractal} for the supergranular fractal dimension with the SOHO dopplergram data. The slight difference of about 4\% can be attributed to the differences in the data used in that study (SoHO instead of KSO) and the differing time-frames in the two studies.

\begin{table}
	\centering
	\begin{tabular}{|c|c|c|}
		\hline
		Cycle phases & Fractal dimension & Goodness of fit  \\
		\hline
		Minimum & 1.47 $\pm$ 0.02 & 0.99\\
		\hline
		Intermediate & 1.3 $\pm$ 0.01 & 0.98\\
		\hline
		Peak & 1.2$\pm$ 0.01  & 0.99\\
		\hline 
	\end{tabular}
\caption{Fractal dimension with different phases of the Solar activity cycle.}
\label{tab:phases}
\end{table}

Our main results are summarized in Table ~\ref{tab:phases}. This shows that the fractal dimension of supergranular cells varies with respect to different phases across the solar cycle. 
In particular, there is a marked trend for the fractal dimension to fall with the level of activity (cf. Figure \ref{fig:A-P}). The goodness of fit is quantified by the linear regression $R^2$ measure.
The low scatter in the data (as reflected in the high goodness of fit) suggests that the phase dependence is unlikely to be an artefact. All the same, a more extensive investigation of fractal exponents is necessary, with special attention paid to enhancing the sample size, in order to obtain dimensions that are statistically more significant.

%\section{Discussion and Conclusion}

In this paper, using Ca II K digitized data obtained from Kodaikanal Solar Observatory, we investigated the long-term behaviour of the supergranular fractal dimension over a solar cycle. Our results indicate that the fractal dimension is anti-correlated with magnetic activity, such that the fractal dimension is just below 1.5 for the minimum phase, 1.3 during intermediate phase and 1.2 at cycle peak. \citet{paniveni2005fractal} used only a time window close to the Solar cycle minimum, whereas the sample here spans the cycle, but is confined to low magnetic activity regions. 

Our results are qualitatively consistent with those of \citet{paniveni2011solar} and \citet{chatterjee2017variation}. In particular, the latter report a similar anticorrelation for quiet regions across a number of cycles obtained from KSO. The average over about 3 cycles is summarized in Table \ref{tab:FD}. This agrees with the anticorrelation found in our data, restricted to the smaller cells (Table \ref{tab:phases}), indicating that this behavior is scale-independent.
\color{blue}\begin{table}
		\centering
	\begin{tabular}{|c|c|c|}
		\hline
		Cycle phase & Fractal dimension $D$ & Average cell scale (Mm) \\
		\hline
	Minimum & 1.22 & 25.4\\
	\hline
	Intermediate & 1.19 & 24.9 \\
	\hline
	Peak & 1.15  & 24.5\\
	\hline
	\end{tabular}
\caption{Fractal dimension across different activity phases over all cell scales, adapted from Figure 8 of Ref. \cite{chatterjee2017variation}. The anticorrelation of scale with activity level here is consistent with a similar result reported by \citet{singh1981dependence}.}
\label{tab:FD}
\end{table}
\color{black}
However, our results pertain to cells that are of a scale smaller than average. This suggests that the anticorrelation between fractal dimension and the activity phase holds across different scales for the quiet region. On the other hand, we find that the fractal dimension values we obtain are slightly larger. This difference may potentially be due to differences in the data sets and methods used to extract $D$. However, in a recent work reported elsewhere \citep{sowmya2022}, we have been able to verify this scale-dependent fractal dimension, suggesting that a multifractal is at play, and therefore that the properties of the convection may evince a degree of scale-dependence.

A point worth noting here is the relatively small value of cell size obtained here, which is comparable to that reported by \citet{krishan2002relationship} using the same data analysis method, which preferentially selects relatively smaller cells. Similar small scales are reported by other studies that tend to extract individual cells, e.g., \citep{parnell2009power}, which reports a cell diameter in the range 13-18 Mm, half of the generally quoted cell sizes that are obtained by the autocorrelation method. These observations underscore that the smaller scale is a selection effect due to the kind of method used. Here we point out that this effect can be understood in terms of the relation between cell size and its boundary properties. The visual inspection method picks out only well delineated cells, which can broadly be interpreted as cells with lower fractal dimension.

Thus, the selection effect in our data can be explained if it is the case that cell size and fractal dimension are positively correlated, which has indeed been reported to be the case \citep{srikanth1999studies}. Solar magnetoconvection is acknowledged to be a turbulent phenomenon \citep{krishan2002relationship}, which suggests that small fluctations in the initial central upflow of the cell can translate to large irregularities of the cell boundary. As a result, the boundaries of larger cells can be expected to be more irregular. Smaller cells are also expected to manifest boundary irregularities, but to a relatively lesser extent. Thus, our requirement for cell boundaries to be regular or well-defined is weighted towards smaller cells. This interpretation could be tested over a larger sample of cells potentially using a machine learning based approach that 

\section{Discussion and conclusions}
An iso-surface has a fractal dimension given by  $D_I$ =(Euclidean dimension)- 1/2 (exponent of the variance)  proposed by \citep{mandelbrot1975geometry} . Thus $D_T =2 –(1/2\times 2/3) =5/3\approx 1.67$ for an isotherm, considering a two-dimensional surface of supergranulation. On the other hand, the pressure variance $\langle p^2\rangle$ is proportional to the square of the velocity variance i.e. $\langle p^2\rangle\propto r^{4/3}$ \cite{batchelor1953theory} and hence for an isobar it is $D_p = 2 - (1/2\times4/3) = 4/3\approx 1.33$. Considering the entire solar cycle, our analysis gives an averaged fractal dimension value closer to $\frac{4}{3}$ than to $\frac{5}{3}$ (cf. Table \ref{tab:phases}), which indicates that the supergranular network is closer to being isobaric than isothermal. This is consistent also with the fractal dimension data derived by \citet{chatterjee2017variation} for supergranulation in both QRs and ARs. Our fractal dimension data is consistent with a turbulent origin of the supergranules.

The proximity of the $D$ values that we find to that corresponding to the isobar limit may be accounted for by noting that the  chromospheric network is situated at the boundary of supergranules. Assuming that supergranules are convective cells, they are expected in mixing-length theory to remain in a full pressure balance with the ambient plasma. This may also explain the difference between our $D$ values and the fractal dimension reported by \citet{nesme1996fractal}, who investigated regions demarcated by intensity delimiters that correspond to isotherm surfaces.

%\citet{singh1981dependence} and \citet{chatterjee2017variation} report a cycle phase dependence of the Ca II K network, with the cells being slightly smaller at the maximum than at the minimum.  then one expects a 5\% reduction in scale with a 15\% increase in the magnetic field strength \citep{chandrasekhar2013hydrodynamic}. Such an increase is not infrequent 

Magnetic fields have the constricting property, by which charged particles cannot cut across magnetic field lines but are constrained to spiral along field lines, essentially as a consequence of the Lorentz force law $\vec{F}_{\rm Lorentz} = \frac{q}{c} {\bf v} \times {\bf B}$.  As a consequence, the field lines are ``frozen in'' with the plasma. In the limit of very high electric conductivity, plasma flow across a magnetic field lines is prohibited as it would create extremely large eddy currents \citep{alfven1942existence}. If supergranular cells correspond to convective cells, then this magnetohydrodynamic effect explains why magnetic fields are swept to the network boundaries. The  effect can also be expected to smoothen the Ca II K cell boundary by supressing fluctuations causing any deviation from the underlying supergranular network boundary, thereby leading to a diminished fractal dimension. Accordingly, during the maximum phase, when there is a rapid dispersal of magnetically active regions, this effect should be more pronounced, leading to a lower value of $D$. Correspondingly, there would be relatively higher ruggedness of the network boundaries during the magnetically quiescent phase, leading to a higher value of $D$. This provides a qualitative explanation for the cycle dependence of the fractal dimension that we report in Table \ref{tab:phases}.

\section{ACKNOWLEDGEMENT}
We thank Professor Jagdev singh IIA, Bengaluru for providing the Kodaikanal solar data. RG and SGM thank Fiaz of IIA for the technical support.
\bibliographystyle{apalike}
\bibliography{./rajani}

\begin{thebibliography}{}

\bibitem[Alfv{\'e}n, 1942]{alfven1942existence}
Alfv{\'e}n, H. (1942).
\newblock Existence of electromagnetic-hydrodynamic waves.
\newblock {\em Nature}, 150(3805):405--406.

\bibitem[Batchelor, 1953]{batchelor1953theory}
Batchelor, G.~K. (1953).
\newblock {\em The theory of homogeneous turbulence}.
\newblock Cambridge university press.

\bibitem[Cannon, 1984]{cannon1984fractal}
Cannon, J. (1984).
\newblock The fractal geometry of nature. by benoit b. mandelbrot.
\newblock {\em The American Mathematical Monthly}, 91(9):594--598.

\bibitem[Chatterjee et~al., 2017]{chatterjee2017variation}
Chatterjee, S., Mandal, S., and Banerjee, D. (2017).
\newblock Variation of supergranule parameters with solar cycles: results from
  century-long kodaikanal digitized ca ii k data.
\newblock {\em The Astrophysical Journal}, 841(2):70.

\bibitem[De~Rosa et~al., 2000]{de2000near}
De~Rosa, M., Duvall, T., and Toomre, J. (2000).
\newblock Near-surface flow fields deduced using correlation tracking and
  time-distance analyses.
\newblock {\em Solar Physics}, 192(1):351--361.

\bibitem[DeRosa and Toomre, 2004]{derosa2004evolution}
DeRosa, M.~L. and Toomre, J. (2004).
\newblock Evolution of solar supergranulation.
\newblock {\em The Astrophysical Journal}, 616(2):1242.

\bibitem[Hathaway et~al., 2000]{hathaway2000photospheric}
Hathaway, D.~H., Beck, J., Bogart, R., Bachmann, K., Khatri, G., Petitto, J.,
  Han, S., and Raymond, J. (2000).
\newblock The photospheric convection spectrum.
\newblock {\em Solar Physics}, 193(1):299--312.

\bibitem[Hirzberger et~al., 2008]{hirzberger2008structure}
Hirzberger, J., Gizon, L., Solanki, S.~K., and Duvall, T.~L. (2008).
\newblock Structure and evolution of supergranulation from local
  helioseismology.
\newblock In {\em Helioseismology, Asteroseismology, and MHD Connections},
  pages 415--435. Springer.

\bibitem[Krishan et~al., 2002]{krishan2002relationship}
Krishan, V., Paniveni, U., Singh, J., and Srikanth, R. (2002).
\newblock Relationship between horizontal flow velocity and cell size for
  supergranulation using soho dopplergrams.
\newblock {\em Monthly Notices of the Royal Astronomical Society},
  334(1):230--232.

\bibitem[Mandal et~al., 2017]{mandal2017association}
Mandal, S., Chatterjee, S., and Banerjee, D. (2017).
\newblock Association of supergranule mean scales with solar cycle strengths
  and total solar irradiance.
\newblock {\em The Astrophysical Journal}, 844(1):24.

\bibitem[Mandelbrot, 1975]{mandelbrot1975geometry}
Mandelbrot, B.~B. (1975).
\newblock On the geometry of homogeneous turbulence, with stress on the fractal
  dimension of the iso-surfaces of scalars.
\newblock {\em Journal of Fluid Mechanics}, 72(3):401--416.

\bibitem[Meunier, 1999]{meunier1999fractal}
Meunier, N. (1999).
\newblock Fractal analysis of michelson doppler imager magnetograms: a
  contribution to the study of the formation of solar active regions.
\newblock {\em The Astrophysical Journal}, 515(2):801.

\bibitem[Meunier et~al., 2008]{meunier2008supergranules}
Meunier, N., Roudier, T., and Rieutord, M. (2008).
\newblock Supergranules over the solar cycle.
\newblock {\em Astronomy \& Astrophysics}, 488(3):1109--1115.

\bibitem[Meunier et~al., 2007]{meunier2007intensity}
Meunier, N., Tkaczuk, R., and Roudier, T. (2007).
\newblock Intensity variations inside supergranules.
\newblock {\em Astronomy \& Astrophysics}, 463(2):745--753.

\bibitem[Muller et~al., 1987]{muller1987dynamics}
Muller, R., Roudier, T., Malherbe, J., and Mein, P. (1987).
\newblock Dynamics of the solar granulation.
\newblock {\em Publications of the Astronomical Institute of the Czechoslovak
  Academy of Sciences}, 1:175.

\bibitem[Nesme-Ribes et~al., 1996]{nesme1996fractal}
Nesme-Ribes, E., Meunier, N., and Collin, B. (1996).
\newblock Fractal analysis of magnetic patterns from meudon spectroheliograms.
\newblock {\em Astronomy and Astrophysics}, 308:213--218.

\bibitem[Paniveni, 2015]{paniveni2015supergranulation}
Paniveni, U. (2015).
\newblock Supergranulation, a convective phenomenon.
\newblock {\em IAU General Assembly}, 29:2146218.

\bibitem[Paniveni et~al., 2004]{paniveni2004relationship}
Paniveni, U., Krishan, V., Singh, J., and Srikanth, R. (2004).
\newblock Relationship between horizontal flow velocity and cell lifetime for
  supergranulation from soho dopplergrams.
\newblock {\em Monthly Notices of the Royal Astronomical Society},
  347(4):1279--1281.

\bibitem[Paniveni et~al., 2005]{paniveni2005fractal}
Paniveni, U., Krishan, V., Singh, J., and Srikanth, R. (2005).
\newblock On the fractal structure of solar supergranulation.
\newblock {\em Solar Physics}, 231(1):1--10.

\bibitem[Paniveni et~al., 2010]{paniveni2010activity}
Paniveni, U., Krishan, V., Singh, J., and Srikanth, R. (2010).
\newblock Activity dependence of solar supergranular fractal dimension.
\newblock {\em Monthly Notices of the Royal Astronomical Society},
  402(1):424--428.

\bibitem[Paniveni et~al., 2011]{paniveni2011solar}
Paniveni, U., Krishan, V., Singh, J., and Srikanth, R. (2011).
\newblock Solar cycle phase dependence of supergranular fractal dimension.
\newblock {\em Journal of Astrophysics and Astronomy}, 32(1-2):265.

\bibitem[Parnell et~al., 2009]{parnell2009power}
Parnell, C., DeForest, C., Hagenaar, H., Johnston, B., Lamb, D., and Welsch, B.
  (2009).
\newblock A power-law distribution of solar magnetic fields over more than five
  decades in flux.
\newblock {\em The Astrophysical Journal}, 698(1):75.

\bibitem[Raju et~al., 1998]{raju1998dependence}
Raju, K., Srikanth, R., and Singh, J. (1998).
\newblock The dependence of chromospheric ca ii k network cell sizes on solar
  latitude.
\newblock {\em Solar Physics}, 180(1):47--51.

\bibitem[Rieutord et~al., 2008]{rieutord2008solar}
Rieutord, M., Meunier, N., Roudier, T., Rondi, S., Beigbeder, F., and Pares, L.
  (2008).
\newblock Solar supergranulation revealed by granule tracking.
\newblock {\em Astronomy \& Astrophysics}, 479(1):L17--L20.

\bibitem[Rieutord and Rincon, 2010]{rieutord2010sun}
Rieutord, M. and Rincon, F. (2010).
\newblock The sun’s supergranulation.
\newblock {\em Living Reviews in Solar Physics}, 7(1):1--82.

\bibitem[Rincon and Rieutord, 2018]{rincon2018sun}
Rincon, F. and Rieutord, M. (2018).
\newblock The sun’s supergranulation.
\newblock {\em Living Reviews in Solar Physics}, 15(1):1--74.

\bibitem[Singh and Bappu, 1981]{singh1981dependence}
Singh, J. and Bappu, M. (1981).
\newblock A dependence on solar cycle of the size of the ca+ network.
\newblock {\em Solar Physics}, 71(1):161--168.

\bibitem[Sowmya et~al., 2022]{sowmya2022}
Sowmya, G.~M., Rajani, G., Paniveni, U., and Srikanth, R. (2022).
\newblock Supergranular fractal dimension and solar rotation.
\newblock Under submission.

\bibitem[Srikanth, 1999]{srikanth1999studies}
Srikanth, R. (1999).
\newblock {\em Studies on the Solar Chromospheric Network}.
\newblock PhD thesis, Indian Institute of Science, Bangalore.

\bibitem[Srikanth et~al., 1999]{srikanth1999chromospheric}
Srikanth, R., Raju, K., and Singh, J. (1999).
\newblock The chromospheric network: Dependence of cell lifetime on
  length-scale.
\newblock {\em Solar Physics}, 184(2):267--280.

\end{thebibliography}
\end{document}